\documentclass[a4paper,fleqn,usenatbib]{mnras}

\usepackage[T1]{fontenc}
\usepackage{ae,aecompl}
\usepackage{graphicx}	
\usepackage{amsmath}	
\usepackage{amssymb}	
\usepackage{gnuplot-lua-tikz}

\newcommand{\nugc}{{\scriptsize$\nu^{2}$GC}}
\defcitealias{nu2gc}{M16}
\defcitealias{GS15}{GS15}

\title[Effect of the seed black hole mass]{Theoretical reevaluations of the black hole mass -- bulge mass relation - I. Effect of the seed black hole mass}

\author[H.~Shirakata et al.]
  {Hikari~Shirakata,$^1$\thanks{E-mail: shirakata@astro1.sci.hokudai.ac.jp} Toshihiro~Kawaguchi,$^2$ Takashi~Okamoto, $^{1}$ Ryu~Makiya,$^{3,4}$ 
    \newauthor Tomoaki~Ishiyama,$^{5}$ Yoshiki Matsuoka,$^{6,7}$ Masahiro~Nagashima,$^8$ Motohiro~Enoki,$^9$ 
    \newauthor Taira~Oogi,$^3$ and Masakazu~A.~R.~Kobayashi$^{10}$\\
  $^1$Department of Cosmosciences, Hokkaido University, N10 W8, Kitaku, Sapporo, 060-0810, Japan\\
  $^2$Department of Liberal Arts and Sciences, Sapporo Medical University, S1W17, Chuo-ku, Sapporo, 060-8556, Japan\\
  $^3$Kavli Institute for the Physics and Mathematics of the universe, Todai Institutes for Advanced Study, the University of Tokyo, Kashiwa, 277-8583 Japan\\
  $^4$Max-Planck-Institut fur Astrophysik, Karl-Schwarzschild Str. 1, D-85741 Garching, Germany\\
  $^5$Institute of Management and Information Technologies, Chiba University, 1-33, Yayoi-cho, Inage-ku, Chiba, 263-8522, Japan\\
  $^6$National Astronomical Observatory of Japan, Mitaka, Tokyo 181-8588, Japan\\
  $^7$Department of Astronomy, School of Science, Graduate University for Advanced Studies, Mitaka, Tokyo 181-8588, Japan\\
  $^8$Faculty of Education, Bunkyo University, Koshigaya, Saitama 343-8511, Japan\\
  $^9$Faculty of Business Administration, Tokyo Keizai University, Kokubunji, Tokyo, 185-8502, Japan\\
  $^{10}$Faculty of Natural Sciences, National Institute of Technology, Kure College, 2-2-11, Agaminami, Kure, Hiroshima, 737-8506, Japan
  }
\date{Accepted XXX. Received YYY; in original form ZZZ}

\pubyear{2016}

\begin{document}
\label{firstpage}
\pagerange{\pageref{firstpage}--\pageref{lastpage}}
\maketitle

\begin{abstract}
We explore the effect of varying the mass of the seed black hole 
on the resulting black hole mass -- bulge mass relation at  $z \sim 0$,  
using a semi-analytic model of galaxy formation combined with large cosmological $N$-body simulations.
We constrain our model by requiring the observed properties of galaxies at  $z \sim 0$ are reproduced. 
In keeping with previous semi-analytic models, we place a seed black hole immediately after a galaxy forms. 
When the mass of the seed is set at $10^5 M_\odot$, 
we find that the model results become inconsistent with recent observational results 
of the black hole mass -- bulge mass relation for dwarf galaxies. 
In particular, the model predicts that bulges with $\sim 10^9 M_\odot$ 
harbour larger black holes than observed. 
On the other hand, when we employ seed black holes with $10^3 M_\odot$, 
or randomly select their mass within a $10^{3-5} M_\odot$ range, 
the resulting relation is consistent with observation estimates, including the observed dispersion. 
We find that to obtain stronger constraints on the mass of seed black holes, 
observations of less massive bulges at $z \sim 0$ are a more powerful comparison 
than the relations at higher redshifts.

\end{abstract}




\section{Introduction}
Observations have found luminous quasars at $z > 6$,
with masses of supermassive black holes (SMBHs) estimated at $\sim 10^{9} M_\odot$ \citep{Mortlock11, Wu15}.
Larger SMBH masses at higher redshifts require either
(i) relatively heavier seed black holes (BHs) with $\sim 10^5 M_\odot$
\citep[e.g.,][]{LN06}, or
(i\hspace{-.1em}\nobreak i) super-Eddington accretion for rapid growth of BHs
\citep{Rees92, Kawaguchi03}.
Both these mechanisms are potentially possible: 
(i) The massive seed BHs 
can be formed as the end products of gas collapse 
with virial temperatures $\geq 10^4$ K without molecular cooling \citep{BVR06}.
(i\hspace{-.1em}i) Estimations of accretion rates 
and duration of the super-Eddington accreting active galactic nuclei (AGNs),
together with the observed trend of higher Eddington ratios
at higher redshift \citep[e.g.,][]{MD04,Nobuta12},
indicate that BHs have grown via super-Eddington accretion
in early universe \citep{Kawaguchi04June}.
These two mechanisms have been actively discussed. 

Because of the uncertainty of the BH formation mechanism 
and limited computational resources, 
most cosmological hydrodynamic simulations and
semi-analytic models of galaxy formation (hereafter SA models)
have treated the formation of BHs 
by putting a seed BH with a set mass at the centre of each galaxy.
For example, in \cite{Barber16}, who use the EAGLE simulation,
a seed BH of $10^5 h^{-1} M_\odot$ is placed
by converting the bound gas particle with the highest density
when a collapsed halo with $\geq 10^{10}h^{-1}M_\odot$ contains no BHs.
\cite{Okamoto08M}, on the other hand, employ a seed BH mass
with $10^{2} M_\odot$ and place the seed at the stellar density maxima
when a virialized dark matter halo that consists of more than
1000 dark matter particles does not contain any BHs.
Other cosmological simulations \citep[e.g.,][]{BS09} assume that
the seed BH mass is proportional to the gas mass in the host halo.
SA models treat the birth of BHs in mainly two different ways.
In the first, a central BH is born 
when a galaxy initially merges with other galaxies \citep[e.g.,][]{KH00,Enoki03,Malbon07,Lagos09}.
In these models, the initial BH mass depends on the amount of the cold gas in the merging system.
In the second method, a seed BH of a fixed mass is placed immediately after a galaxy forms: 
$10^{2} M_\odot$ \citep{Menci03}, $10^{3} M_\odot$ \citep{FMS15},
or $10^{5} M_\odot$ \citep{nu2gc}.
For a deeper understanding of seed BHs from SA models, 
\cite{PVS16} and \cite{Valiante16J} focus on BH growth 
only in early universe ($z \gtrsim 5$) and suggest that  
100 $M_\odot$ seed BHs at $z \gtrsim 23$ accretes gas via
major mergers at super-Eddington rates, forming
$10^5 M_\odot$ BHs at $z \sim 17$.
The mass of the seed BHs has previously been presumed to have only a small impact on the statistical properties
of galaxies, AGNs and SMBHs, unless the mass accretion rate depends on black hole mass
and a large amount of gas from their host galaxies is accreted by the BHs.

Many observations \citep[e.g.,][]{KR95, HR04, MM13, Magorrian98}
have suggested that 
the mass of SMBHs ($M_\mathrm{BH}$) correlates with
the stellar mass of their host bulges ($M_\mathrm{bulge}$)
at $z \sim 0$ (hereafter the $M_\mathrm{BH}$ -- $M_\mathrm{bulge}$ relation).
They indicate a nearly proportional $M_\mathrm{BH}$ -- $M_\mathrm{bulge}$ relation.
The $M_\mathrm{BH}$ -- $M_\mathrm{bulge}$ relation may suggest
that SMBHs have co-evolved with their host galaxies.
This co-evolution of BH and host galaxies have motivated
theoretical studies investigating the mechanism of BH feeding, AGN activities,
and the energetic feedback from AGNs in the context of the galaxy formation
\citep[see, however, ][]{JM11}.

It is worth noting that recent observational studies 
\citep[e.g.,][]{Graham12F, SGS13} have suggested
that for $M_\mathrm{bulge} \lesssim 10^{10} M_\odot$,
the $M_\mathrm{BH}$ -- $M_\mathrm{bulge}$ relation becomes
quadratic instead of the linear relation
found for more massive bulges.
It has also suggested that this quadratic relation continues
down to $M_\mathrm{BH} \sim 10^{5} M_\odot$.

In this {\it Letter}, we investigate whether the mass of seed BHs
affects model predictions of the local $M_\mathrm{BH}$ -- $M_\mathrm{bulge}$ relation
at $M_\mathrm{bulge} \lesssim 10^{10} M_\odot$
by using an SA model.
In Section 2 we briefly review our models regarding the growth of SMBHs.
In Section 3 we present our results. 
We summarize and discuss the results in Section 4.

\section{Methods}
\label{sec:methods} 
We employ a revised version of the SA model,  
``{\it New Numerical Galaxy Catalogue''} (\nugc; \citealt[][hereafter M16]{nu2gc}),
where the SMBH growth and AGN properties are summarized by
\cite{Enoki03}, \cite{Enoki14}, and \cite{Shirakata15}.
The revised model differs from \citetalias{nu2gc} in the following points;
(i) the model of merger driven spheroid formation and 
(i\hspace{-.1em}\nobreak i) the growth of spheroids via disc instability. 
Further details are given in Sec. 2.1
Despite these changes, we will show later that 
the main result of this {\it Letter}
remains unchanged even if we use the \citetalias{nu2gc} model.

We create merging histories of dark matter haloes from
state-of-the-art cosmological $N$-body simulations \citep{Ishiyama15}, 
which have a high mass resolution and large volume compared to previous simulations
(e.g., 4 times better mass resolution
compared with Millennium simulations, \citealt{Springel05June}).
Table \ref{tab:N-body} summarizes basic properties of the simulations used in this {\it Letter}.
The \nugc~-L simulation has $16^3$ times larger volume 
and the same mass resolution compared with the \nugc~-SS simulation.
When calculating the SMBH mass function at $z \sim 6$, we use the \nugc~-L
since SMBHs in the early universe are rare.
The \nugc~-H2 simulation has the same box size and $\sim 64$ times higher mass resolution
compared with the \nugc~-SS simulation.
Since we are interested in small galaxies in this {\it Letter},
we employ the \nugc~-H2 simulation in most cases, while
the \nugc~-SS simulation is used for resolution studies.
The details of the merger trees are given in \cite{Ishiyama15}.
We consider mergers of galaxies, atomic gas cooling, gas heating by UV feedback
and feedbacks via supernovae and AGNs, and the growth of SMBHs by coalescence and gas fueling.
More detailed descriptions are available in \citetalias{nu2gc}.
Throughout this {\it Letter}, we assume a $\Lambda$CDM universe with
the following parameters:
$\Omega_{0}=0.31$,  $\lambda_{0}=0.69$, 
$\Omega_\mathrm{b} = 0.048$, 
$\sigma_8 = 0.83$, $n_\mathrm{s} = 0.96$, and a Hubble constant of 
$H_0 = 100~h~\mathrm{km}~\mathrm{s}^{-1}$~Mpc$^{-1}$, where $h=0.68$ 
\citep{Planck}.
\begin{table*}
\centering
  \begin{tabular}{lllccc}
  \hline
    Name & $N$ & $L$~[$h^{-1}$ Mpc] & $m~[h^{-1} M_\odot]$ & $M_\mathrm{min}~[h^{-1} M_\odot]$ & $M_\mathrm{max}~[h^{-1} M_\odot]$ \\
      \hline
    \nugc {\scriptsize -L} & $8192^3$ & 1120.0 & $2.20\times 10^8 $ & $8.79\times 10^9$ & $4.11\times 10^{15}$ \\
    \nugc {\scriptsize -SS} & $512^3$ & 70.0 & $2.20\times 10^8 $ & $8.79\times 10^9$ & $6.58\times 10^{14}$ \\
    \nugc {\scriptsize -H2} & $2048^3$ & 70.0 & $3.44\times 10^6 $ & $1.37\times 10^8$ & $4.00\times 10^{14}$ \\
    \hline
  \end{tabular}
\caption{Properties of the \nugc~simulations. $N$ is the number of simulated particles, $L$ is the comoving box size,
  $m$ is the individual mass of a dark matter particle, $M_\mathrm{min}$ is the mass of the smallest haloes ($= 40\times m$) which corresponds to
  the mass resolution, and $M_\mathrm{max}$ is the mass of the largest halo in each simulation.}
\label{tab:N-body}
\end{table*}

\subsection{Spheroid formation}
\label{sec:spheroid}
We assume that the spheroid within galaxies grows
via starbursts and the migration of disc stars.
In the \citetalias{nu2gc} model, 
we considered starbursts only triggered in the major mergers of galaxies.
Such a model, however, cannot reproduce AGN luminosity functions at low and high redshifts simultaneously
\citep[e.g.,][]{Enoki14}.
In the revised version of the model in \nugc, 
we therefore assume that starbursts are triggered not only by
major mergers of galaxies but also by minor mergers and disc instability.

For mergers, we introduce the model of merger driven-spheroid
formation proposed by \cite{Hopkins09F} based on hydrodynamic simulations,
similar to the SA model by \cite{Somerville15}.
After dark matter haloes merge together, we regard the central galaxy in the most massive 
progenitor halo as the new central galaxy in the combined halo.
We next consider a merger of two galaxies 
\footnote{ We consider galaxy mergers via dynamical friction (central-satellite mergers)
           and via random collisions (satellite-satellite merger) 
           without following subhalo orbits.
           We use the merging rate by the random collision estimated by \cite{MH97}.
           The detailed descriptions of these mergers are found in Sec. 2.5 in \citetalias{nu2gc}.}
where the primary galaxy has a larger baryon mass (cold gas $+$ stars)
than a secondary galaxy.
In all such merger events, we assume that all stars in the secondary are absorbed in the bulge of the primary,
and a part of the disc (cold gas $+$ stars) of the primary is also turned into the bulge component.
We assume that cold gas that falls into the primary's bulge
is consumed by a starburst.
Similarly to \citetalias{nu2gc}, we assume that the starburst
and the gas accretion onto the SMBH start at the same time.
The beginning time is assigned randomly within the timestep.

We also introduce the spheroid formation by disc instability
in order to form SMBHs at $z \sim 6$. 
The spheroids formed through this mechanism might become so-called `pseudo bulges',
although we do not differentiate bulges formed by this process 
from those formed by mergers.
The inclusion of disc instabilities are important as 
strongly self-gravitating galactic discs are likely to be dynamically unstable.
Moreover, without the disc instability, 
we cannot form high redshift ($z \sim 6$) SMBHs with
$M_\mathrm{BH} \gtrsim 10^9 M_\odot$.

Following \cite{MMW98} and \cite{Cole00},
a galactic disc becomes unstable when
\begin{equation}
  f_\mathrm{DI} \equiv \frac{V_\mathrm{d}}{(GM_\mathrm{disc}/r_\mathrm{disc})^{1/2}} < f_\mathrm{DI, crit},
\end{equation}
where $G$ is the gravitational constant, $V_\mathrm{d}$, $M_\mathrm{disc}$, and $r_\mathrm{d}$ 
are the rotation velocity, the mass of cold gas $+$ stars, and the half mass radius of the galactic disc, respectively. 
Since $f_\mathrm{DI,crit}$ depends on the gas fraction and density profile of a galactic disc
\citep[e.g.,][]{Efstathiou82,CST95},
we consider $f_\mathrm{DI,crit}$ as an adjustable parameter, 
whose value is chosen so as to reproduce the observed stellar mass function of bulges at $z \sim 0$.
We assume that an unstable disc is completely destroyed and reforms into a bulge with a starburst.

\subsection{Growth of SMBHs}
\label{sec:SMBH}
When galaxies merge with each other or the galactic disc becomes dynamically unstable,
a starburst occurs and a small fraction of cold gas gets accreted by the SMBH via the following relation:
\begin{equation}
  M_\mathrm{acc} = f_\mathrm{BH}~~\Delta M_\mathrm{*,burst},
\end{equation}
where $M_\mathrm{acc}$ is the accreted cold gas mass which is proportional to
the stellar mass newly formed by a starburst, $\Delta M_\mathrm{*,burst}$.
In this {\it Letter}, we set $f_\mathrm{BH} = 0.01$
and assume that the accretion rate onto an SMBH is not limited by the Eddington accretion rate.

SMBHs also grow when BH-BH coalesce as their host galaxies merge.
The timescale of the coalescence is difficult to determine.
Similar to earlier work \citep[e.g.,][]{KH00, Enoki03, Somerville08},
we thus assume BHs merge instantaneously 
when their host galaxies merge.

We also consider radio-mode AGN feedback process as in \cite{Bower06}.
Namely, when the halo cooling time becomes
longer than the free-fall time at the cooling radius, and the cooling luminosity
becomes smaller than the accretion luminosity,
a hot halo gas is prevented from cooling.

\subsection{Seed Black Holes}
\label{sec:seed}
We place a seed BH immediately after a galaxy forms.
In this {\it Letter}, 
we present results with $M_\mathrm{BH,seed} = 10^3 M_\odot$
(hereafter `light seed model') where $M_\mathrm{BH,seed}$ is the seed BH mass, 
  and $10^5 M_\odot$ (`massive seed model').
We also test the case 
in which $M_\mathrm{BH,seed}$ takes uniformly random values
in the logarithmic scale in the 
range of $3 \leq \log(M_\mathrm{BH,seed}/M_\odot) \leq 5$
(hereafter `random seed model').

A galaxy is born when hot gas in the host halo 
cools efficiently.
The hot gas cools by atomic cooling when
the virial temperature, $T_\mathrm{vir}$, is larger than $10^{4}$ K.
We also consider the heating effect by the UV background.
We employ fitting formulae of the characteristic halo mass, $M_c (z)$, 
obtained from \cite{Okamoto08N},
below which haloes become baryon deficient.
We find that runs \nugc~-SS and -L do not resolve haloes with
$M_c (z)$ in any redshift, while the \nugc~-H2 does resolve haloes with $M_c (z)$ 
at $z \lesssim 5$ and those with $T_\mathrm{vir} < 10^4$ K at $z \lesssim 3$.
This difference, however, does not affect the main conclusion
of this {\it Letter} as we will show later.

\section{Results}
\label{sec:results}
In this section we present the effect of the seed BH mass
on the $M_\mathrm{BH}-M_\mathrm{bulge}$ relation at $z \sim 0$.
Our model reproduces the observed galaxy luminosity functions, HI,
BH, and bulge stellar mass functions at $z \sim 0$.
Values of the model parameters used in this {\it Letter} are listed in Table~\ref{table:parameters}.
The description of the parameters is given in \citetalias{nu2gc}. 
\begin{table}
\begin{center}
    \begin{tabular}{ll | ll | ll}
    \hline
      $\alpha_\mathrm{star}$ & -2.131 & $\tau_\mathrm{V0}$ & $2.5\times 10^{-9}$ & $\epsilon_\mathrm{SMBH}$ & 0.310  \\  
      $\epsilon_\mathrm{star}$ & 0.203 & $f_\mathrm{mrg}$ & 1.0 & $\alpha_\mathrm{cool}$ & 9.510 \\
      $\alpha_\mathrm{hot}$ & 4.080 & $\kappa_\mathrm{diss}$ & 2.0 & $f_\mathrm{DI,crit}$ & 1.1 \\
      $V_\mathrm{hot}$ &  129.7 [km/s] & $f_\mathrm{BH}$ & 0.01 \\
    \hline
    \end{tabular}
  \caption{Values of model parameters in this {\it Letter}.}
  \label{table:parameters}
\end{center}
\end{table}

\begin{figure}
\centering
  \scalebox{0.5}{
    \input{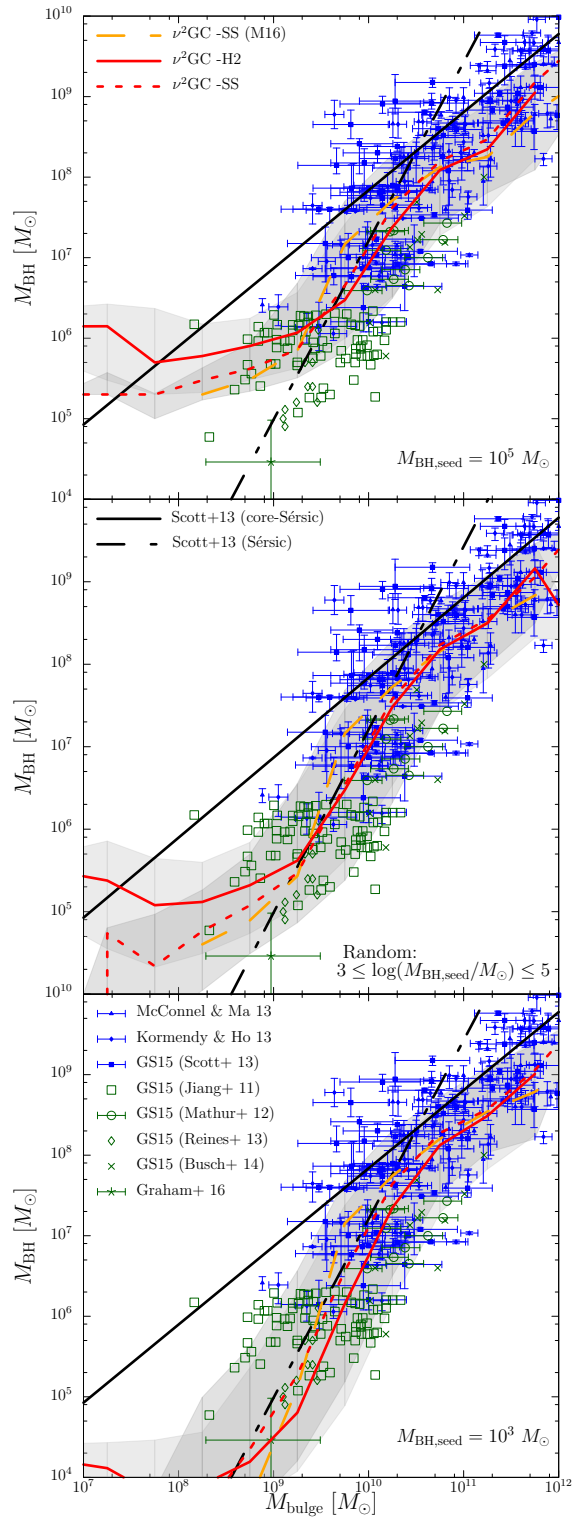}
  }
  \caption{$M_\mathrm{BH}$ -- $M_\mathrm{bulge}$ relations at $z \sim 0$
           for different $M_\mathrm{BH,seed}$; 
           the massive (top), random (middle), and light (bottom) seed models.
           Red dashed and solid lines present the results
           of the \nugc~-SS and -H2 simulations, respectively.
           Red lines track the median, and shaded regions indicate 10-90 percentile of the models.
           Orange long dashed lines present the result of \protect\citetalias{nu2gc}
           with the \nugc~-SS simulation.
           Blue filled symbols indicate observational results obtained from
           \protect\cite{MM13}, \protect\cite{KH13_review}, and
           \protect\citetalias{GS15}\protect\footnotemark
           (triangles, diamonds, and squares, respectively).
           Green open symbols are AGN sample obtained from
           \protect\citetalias{GS15},(see the text 
           for more details).
           Green asterisks correspond LEDA 87300 \protect\citep{GCS16, Baldassare15}.
           Black dot-dashed and solid lines depict the scaling relations \protect\citep{SGS13}.
           Since the $M_\mathrm{BH}$ -- $M_\mathrm{bulge}$ relation is
           sensitive to the mass of seed BHs,
           most seed BHs should not set to $10^5 M_\odot$ for reproducing 
           the observed local relation.}
  \label{fig:main}
\end{figure}

In Fig.~\ref{fig:main}, we present the $M_\mathrm{BH}-M_\mathrm{bulge}$
relation at $z \sim 0$ predicted by the massive seed model (top panel) and light seed model (bottom panel).
For the observational data with $M_\mathrm{BH} \lesssim 10^6 M_\odot$,
we use the data obtained from \cite{GS15} (hearafter GS15).
This work re-estimated the bulge and BH masses obtained by previous work
\citep{Jiang11, Mathur12, SGS13, RGG13,Busch14}.
We also plot LEDA 87300 whose BH mass is originally estimated by \cite{Baldassare15}
and re-evaluated by \cite{GCS16}.
Almost all of the observational samples with $M_\mathrm{BH} \lesssim 10^{6} M_\odot$
have active BHs.
In our model, the $M_\mathrm{BH}$ -- $M_\mathrm{bulge}$ relation does not change
when we only plot AGNs.

Although all of our models reproduce the relation 
at $M_\mathrm{bulge} \gtrsim 10^{10} M_\odot$,
the massive seed model is inconsistent with 
the recent observational estimates for dwarf galaxies with $M_\mathrm{bulge} \lesssim 10^{10} M_\odot$,
in the sense that the predicted BH masses (shaded region) are larger than the observational estimates.
We present the results by the $N$-body simulations with the same box size and
different mass resolution (\nugc~-SS and -H2)
in all panels of Fig. \ref{fig:main}.
We find that the effect of the mass resolution of $N$-body simulations
clearly appears with $M_\mathrm{bulge} \lesssim 10^{9} M_\odot$.
Nonetheless, the mass resolution does not affect 
our conclusion.
Middle panel of Fig. \ref{fig:main} shows the result of the random seed model.
We find that the random seed and light seed models  
reproduce the relation and its scatter well.
\footnotetext{Originally obtained from \protect\cite{SGS13}.}
These two successful models provide the same results
in the range of $M_\mathrm{BH} \gtrsim 10^{5.5} M_\odot$
below which these models have significantly different slope
of the relation.
More observational data with $M_\mathrm{BH} \lesssim 10^{5.5} M_\odot$ are required 
for stronger constraints on the mass distribution of the seed BHs.

The sample of \cite{Jiang11} seems to have the floor in the $M_\mathrm{BH}$ -- $M_\mathrm{bulge}$ relation
at $M_\mathrm{BH} \sim 10^{5-6} M_\odot$.
\cite{Jiang11} thus support the results obtained from the simulations of 
\cite{VN09} with seed BHs of $10^5 M_\odot$.
On the other hand, our model prefers the lower seed BH mass than $10^5 M_\odot$
to explain the $M_\mathrm{BH}$ -- $M_\mathrm{bulge}$ relation obtained from \cite{Jiang11}.

The black dashed and solid lines in Fig \ref{fig:main} depicts
the scaling relations \citep{SGS13}.
  \footnote{\cite{SGS13} classified galaxies by their bulge surface brightness profiles:
    core-S\'ersic galaxies (bulge surface brightness profiles have a partially depleted core)
    and S\'ersic galaxies (bulge surface brightness profiles are well-fitted by a single S\'ersic model).}    
Our models exhibit the slightly lower $M_\mathrm{BH}$ -- $M_\mathrm{bulge}$ relation
than the scaling relation at $M_\mathrm{bulge} \gtrsim 10^{10} M_\odot$.
In this region, $M_\mathrm{bulge}$ evaluated from observations
is potentially biased in favor of larger stellar masses
\citep[e.g.,][]{Shankar16Mar}.

Our models also exhibit the transition of the slope
in the $M_\mathrm{BH}$ -- $M_\mathrm{bulge}$ relations
from quadratic to near-linear.
We also plot $M_\mathrm{BH}$ -- $M_\mathrm{bulge}$ relations
obtained by \citetalias{nu2gc} in Fig. \ref{fig:main}, 
in which starbursts in the bulge and the gas fueling to a BH
are only triggered by major mergers.
They also show the ``bend'', 
meaning that bulge and BH growth via disc instability
has small impact on the ``bend'' of 
the $M_\mathrm{BH}$ -- $M_\mathrm{bulge}$ relation.
We have confirmed that stellar feedback is responsible for
the quadratic relation
as suggested by \cite{FMS15}.

\begin{figure}
  \centering
  \includegraphics[width=\columnwidth]{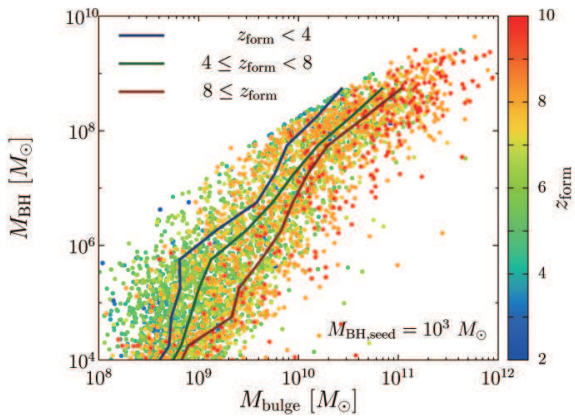}
  \caption{The $M_\mathrm{BH}$ -- $M_\mathrm{bulge}$ relation 
           at $z \sim 0$ with the \nugc~-SS simulation and 
           $M_\mathrm{BH,seed} = 10^3 M_\odot$.
           The color indicates the redshift at which the host galaxies
           newly formed ($z_\mathrm{form}$).
           Solid lines shows the median value for
           different $z_\mathrm{form}$ ranges;
           $z_\mathrm{form} < 4$ (dark brue),
           $4 \leq z_\mathrm{form} < 8$ (dark green), and 
           $8 \leq z_\mathrm{form}$ (dark red).
           For a given SMBH mass, 
           bulges become more massive with higher $z_\mathrm{form}$.}
  \label{fig:z_col}
\end{figure}

Next, we investigate the origin of the scatter
of the $M_\mathrm{BH}$ -- $M_\mathrm{bulge}$ relation.
Fig. \ref{fig:z_col} indicates the distribution of 
the redshift at which galaxies newly form ($z_\mathrm{form}$).
We predict that the scatter of the $M_\mathrm{BH}$ -- $M_\mathrm{bulge}$ relation 
can be related to the difference of $z_\mathrm{form}$.
Three solid lines indicate the relation 
with different ranges of $z_\mathrm{form}$: 
$z_\mathrm{form} < 4$, 
$4 \leq z_\mathrm{form} < 8$, and  
$8 \leq z_\mathrm{form}$.
We find that more massive systems form at higher redshift.
We also find 
that SMBHs become more massive with lower $z_\mathrm{form}$
for a given $M_\mathrm{bulge}$ and 
that the stellar mass of the bulge 
is larger with higher $z_\mathrm{form}$ for a given $M_\mathrm{BH}$.  
These might be because large amount of gas gets accreted by the SMBHs 
immediately after $z_\mathrm{form}$.
Galaxies which formed later are subject to gas-rich processes,
such as major mergers or disc instability.
On the other hand, galaxies hosting the same mass of the SMBHs with higher $z_\mathrm{form}$
have experienced more gas-poor processes of spheroid growth, such as dry mergers,
than lower $z_\mathrm{form}$ conterparts;
their bulges increase their masses without feeding central BHs. 

This trend is inconsistent with  \cite{MFT00} who suggest that
$M_\mathrm{BH}/M_\mathrm{bulge}$ ratios are higher in galaxies
with older stellar age.
We will investigate the origin of the scatter
of $M_\mathrm{BH}$ -- $M_\mathrm{bulge}$ relation in a forthcoming paper.

\section{Summary and Discussion}
\label{sec:Discussion}
We have investigated how the mass of the seed BHs
affects model predictions of the local $M_\mathrm{BH}$ -- $M_\mathrm{bulge}$
relation by using an SA model, \nugc.
We find that seed BHs should not be dominated by those as massive as $10^{5}M_\odot$
to reproduce the observed $M_\mathrm{BH}$ -- $M_\mathrm{bulge}$ relation
at $z \sim 0$ over a wide range of bulge masses down to $M_\mathrm{bulge} \lesssim 10^{10} M_\odot$.
To obtain stronger constraints 
of the mass distribution for the seed BHs, 
observations of $M_\mathrm{BH} \lesssim 10^{5.5} M_\odot$ 
would be required.

The results in this {\it Letter} are consistent
with cosmological hydrodynamic simulations performed by \cite{Angls-Alczar15, Angls-Alczar16}
which suggest that the $M_\mathrm{BH}$ -- $M_\mathrm{bulge}$ relation
converges independently of the seed BH mass at $M_\mathrm{bulge} \gtrsim 10^{10} M_\odot$
while at $M_\mathrm{bulge} \lesssim 10^{10} M_\odot$, seed BH mass becomes important
in the scaling relation.
\cite{Angls-Alczar16} compare BH mass -- galaxy stellar mass relations at $z \sim 0$
with $10^4 h^{-1} M_\odot$ and $10^6 h^{-1} M_\odot$ seed BHs.
They find that in the case with $10^6 h^{-1} M_\odot$ seed BHs,
the relation has a floor which also appears 
in the $M_\mathrm{BH}$ -- $M_\mathrm{bulge}$ relation in our massive seed model.

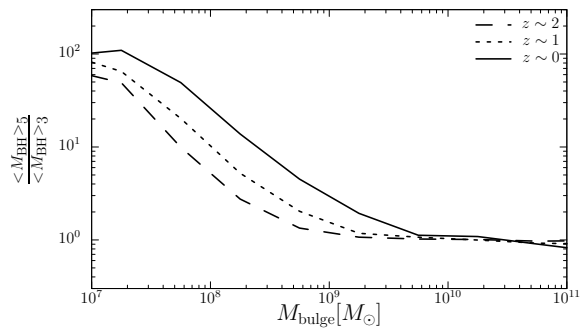
\begin{figure}
\centering
  \scalebox{0.5}{
    \begin{tikzpicture}[gnuplot]
\path (0.000,0.000) rectangle (12.500,8.750);
\gpcolor{color=gp lt color border}
\gpsetlinetype{gp lt border}
\gpsetdashtype{gp dt solid}
\gpsetlinewidth{1.00}
\draw[gp path] (0.000,0.985)--(0.090,0.985);
\draw[gp path] (12.499,0.985)--(12.409,0.985);
\draw[gp path] (0.000,1.293)--(0.090,1.293);
\draw[gp path] (12.499,1.293)--(12.409,1.293);
\draw[gp path] (0.000,1.532)--(0.090,1.532);
\draw[gp path] (12.499,1.532)--(12.409,1.532);
\draw[gp path] (0.000,1.727)--(0.090,1.727);
\draw[gp path] (12.499,1.727)--(12.409,1.727);
\draw[gp path] (0.000,1.892)--(0.090,1.892);
\draw[gp path] (12.499,1.892)--(12.409,1.892);
\draw[gp path] (0.000,2.035)--(0.090,2.035);
\draw[gp path] (12.499,2.035)--(12.409,2.035);
\draw[gp path] (0.000,2.161)--(0.090,2.161);
\draw[gp path] (12.499,2.161)--(12.409,2.161);
\draw[gp path] (0.000,2.274)--(0.180,2.274);
\draw[gp path] (12.499,2.274)--(12.319,2.274);
\node[gp node right] at (-0.184,2.274) {{\Large$10^{0}$}};
\draw[gp path] (0.000,3.016)--(0.090,3.016);
\draw[gp path] (12.499,3.016)--(12.409,3.016);
\draw[gp path] (0.000,3.450)--(0.090,3.450);
\draw[gp path] (12.499,3.450)--(12.409,3.450);
\draw[gp path] (0.000,3.758)--(0.090,3.758);
\draw[gp path] (12.499,3.758)--(12.409,3.758);
\draw[gp path] (0.000,3.997)--(0.090,3.997);
\draw[gp path] (12.499,3.997)--(12.409,3.997);
\draw[gp path] (0.000,4.192)--(0.090,4.192);
\draw[gp path] (12.499,4.192)--(12.409,4.192);
\draw[gp path] (0.000,4.358)--(0.090,4.358);
\draw[gp path] (12.499,4.358)--(12.409,4.358);
\draw[gp path] (0.000,4.500)--(0.090,4.500);
\draw[gp path] (12.499,4.500)--(12.409,4.500);
\draw[gp path] (0.000,4.627)--(0.090,4.627);
\draw[gp path] (12.499,4.627)--(12.409,4.627);
\draw[gp path] (0.000,4.739)--(0.180,4.739);
\draw[gp path] (12.499,4.739)--(12.319,4.739);
\node[gp node right] at (-0.184,4.739) {{\Large$10^{1}$}};
\draw[gp path] (0.000,5.482)--(0.090,5.482);
\draw[gp path] (12.499,5.482)--(12.409,5.482);
\draw[gp path] (0.000,5.916)--(0.090,5.916);
\draw[gp path] (12.499,5.916)--(12.409,5.916);
\draw[gp path] (0.000,6.224)--(0.090,6.224);
\draw[gp path] (12.499,6.224)--(12.409,6.224);
\draw[gp path] (0.000,6.463)--(0.090,6.463);
\draw[gp path] (12.499,6.463)--(12.409,6.463);
\draw[gp path] (0.000,6.658)--(0.090,6.658);
\draw[gp path] (12.499,6.658)--(12.409,6.658);
\draw[gp path] (0.000,6.823)--(0.090,6.823);
\draw[gp path] (12.499,6.823)--(12.409,6.823);
\draw[gp path] (0.000,6.966)--(0.090,6.966);
\draw[gp path] (12.499,6.966)--(12.409,6.966);
\draw[gp path] (0.000,7.092)--(0.090,7.092);
\draw[gp path] (12.499,7.092)--(12.409,7.092);
\draw[gp path] (0.000,7.205)--(0.180,7.205);
\draw[gp path] (12.499,7.205)--(12.319,7.205);
\node[gp node right] at (-0.184,7.205) {{\Large$10^{2}$}};
\draw[gp path] (0.000,7.947)--(0.090,7.947);
\draw[gp path] (12.499,7.947)--(12.409,7.947);
\draw[gp path] (0.000,8.381)--(0.090,8.381);
\draw[gp path] (12.499,8.381)--(12.409,8.381);
\draw[gp path] (0.000,0.985)--(0.000,1.165);
\draw[gp path] (0.000,8.381)--(0.000,8.201);
\node[gp node center] at (0.000,0.677) {{\Large$10^{7}$}};
\draw[gp path] (0.941,0.985)--(0.941,1.075);
\draw[gp path] (0.941,8.381)--(0.941,8.291);
\draw[gp path] (1.491,0.985)--(1.491,1.075);
\draw[gp path] (1.491,8.381)--(1.491,8.291);
\draw[gp path] (1.881,0.985)--(1.881,1.075);
\draw[gp path] (1.881,8.381)--(1.881,8.291);
\draw[gp path] (2.184,0.985)--(2.184,1.075);
\draw[gp path] (2.184,8.381)--(2.184,8.291);
\draw[gp path] (2.432,0.985)--(2.432,1.075);
\draw[gp path] (2.432,8.381)--(2.432,8.291);
\draw[gp path] (2.641,0.985)--(2.641,1.075);
\draw[gp path] (2.641,8.381)--(2.641,8.291);
\draw[gp path] (2.822,0.985)--(2.822,1.075);
\draw[gp path] (2.822,8.381)--(2.822,8.291);
\draw[gp path] (2.982,0.985)--(2.982,1.075);
\draw[gp path] (2.982,8.381)--(2.982,8.291);
\draw[gp path] (3.125,0.985)--(3.125,1.165);
\draw[gp path] (3.125,8.381)--(3.125,8.201);
\node[gp node center] at (3.125,0.677) {{\Large$10^{8}$}};
\draw[gp path] (4.065,0.985)--(4.065,1.075);
\draw[gp path] (4.065,8.381)--(4.065,8.291);
\draw[gp path] (4.616,0.985)--(4.616,1.075);
\draw[gp path] (4.616,8.381)--(4.616,8.291);
\draw[gp path] (5.006,0.985)--(5.006,1.075);
\draw[gp path] (5.006,8.381)--(5.006,8.291);
\draw[gp path] (5.309,0.985)--(5.309,1.075);
\draw[gp path] (5.309,8.381)--(5.309,8.291);
\draw[gp path] (5.556,0.985)--(5.556,1.075);
\draw[gp path] (5.556,8.381)--(5.556,8.291);
\draw[gp path] (5.765,0.985)--(5.765,1.075);
\draw[gp path] (5.765,8.381)--(5.765,8.291);
\draw[gp path] (5.947,0.985)--(5.947,1.075);
\draw[gp path] (5.947,8.381)--(5.947,8.291);
\draw[gp path] (6.107,0.985)--(6.107,1.075);
\draw[gp path] (6.107,8.381)--(6.107,8.291);
\draw[gp path] (6.250,0.985)--(6.250,1.165);
\draw[gp path] (6.250,8.381)--(6.250,8.201);
\node[gp node center] at (6.250,0.677) {{\Large$10^{9}$}};
\draw[gp path] (7.190,0.985)--(7.190,1.075);
\draw[gp path] (7.190,8.381)--(7.190,8.291);
\draw[gp path] (7.740,0.985)--(7.740,1.075);
\draw[gp path] (7.740,8.381)--(7.740,8.291);
\draw[gp path] (8.131,0.985)--(8.131,1.075);
\draw[gp path] (8.131,8.381)--(8.131,8.291);
\draw[gp path] (8.434,0.985)--(8.434,1.075);
\draw[gp path] (8.434,8.381)--(8.434,8.291);
\draw[gp path] (8.681,0.985)--(8.681,1.075);
\draw[gp path] (8.681,8.381)--(8.681,8.291);
\draw[gp path] (8.890,0.985)--(8.890,1.075);
\draw[gp path] (8.890,8.381)--(8.890,8.291);
\draw[gp path] (9.071,0.985)--(9.071,1.075);
\draw[gp path] (9.071,8.381)--(9.071,8.291);
\draw[gp path] (9.231,0.985)--(9.231,1.075);
\draw[gp path] (9.231,8.381)--(9.231,8.291);
\draw[gp path] (9.374,0.985)--(9.374,1.165);
\draw[gp path] (9.374,8.381)--(9.374,8.201);
\node[gp node center] at (9.374,0.677) {{\Large$10^{10}$}};
\draw[gp path] (10.315,0.985)--(10.315,1.075);
\draw[gp path] (10.315,8.381)--(10.315,8.291);
\draw[gp path] (10.865,0.985)--(10.865,1.075);
\draw[gp path] (10.865,8.381)--(10.865,8.291);
\draw[gp path] (11.256,0.985)--(11.256,1.075);
\draw[gp path] (11.256,8.381)--(11.256,8.291);
\draw[gp path] (11.558,0.985)--(11.558,1.075);
\draw[gp path] (11.558,8.381)--(11.558,8.291);
\draw[gp path] (11.806,0.985)--(11.806,1.075);
\draw[gp path] (11.806,8.381)--(11.806,8.291);
\draw[gp path] (12.015,0.985)--(12.015,1.075);
\draw[gp path] (12.015,8.381)--(12.015,8.291);
\draw[gp path] (12.196,0.985)--(12.196,1.075);
\draw[gp path] (12.196,8.381)--(12.196,8.291);
\draw[gp path] (12.356,0.985)--(12.356,1.075);
\draw[gp path] (12.356,8.381)--(12.356,8.291);
\draw[gp path] (12.499,0.985)--(12.499,1.165);
\draw[gp path] (12.499,8.381)--(12.499,8.201);
\node[gp node center] at (12.499,0.677) {{\Large$10^{11}$}};
\draw[gp path] (0.000,8.381)--(0.000,0.985)--(12.499,0.985)--(12.499,8.381)--cycle;
\node[gp node center,rotate=-270] at (-1.558,4.683) {{\huge$\frac{<M_\mathrm{BH}>_{\scriptstyle 5}}{<M_\mathrm{BH}>_{\scriptstyle  3}}$}};
\node[gp node center] at (6.249,0.215) {{\huge$M_\mathrm{bulge} [M_\odot]$}};
\node[gp node left] at (11.211,7.976) {{\Large$z \sim 2$}};
\gpcolor{rgb color={0.000,0.000,0.000}}
\gpsetdashtype{dash pattern=on 10.00*\gpdashlength off 10.00*\gpdashlength }
\gpsetlinewidth{3.00}
\draw[gp path] (10.111,7.976)--(11.027,7.976);
\draw[gp path] (0.000,6.631)--(0.781,6.442)--(2.344,4.732)--(3.906,3.358)--(5.468,2.588)%
  --(7.031,2.347)--(8.593,2.298)--(10.155,2.285)--(11.718,2.242)--(12.499,2.249);
\gpcolor{color=gp lt color border}
\node[gp node left] at (11.211,7.526) {{\Large$z \sim 1$}};
\gpcolor{rgb color={0.000,0.000,0.000}}
\gpsetdashtype{dash pattern=on 2.00*\gpdashlength off 5.00*\gpdashlength }
\draw[gp path] (10.111,7.526)--(11.027,7.526);
\draw[gp path] (0.000,6.988)--(0.781,6.743)--(2.344,5.495)--(3.906,4.044)--(5.468,3.034)%
  --(7.031,2.450)--(8.593,2.346)--(10.155,2.271)--(11.718,2.189)--(12.499,2.176);
\gpcolor{color=gp lt color border}
\node[gp node left] at (11.211,7.076) {{\Large$z \sim 0$}};
\gpcolor{rgb color={0.000,0.000,0.000}}
\gpsetdashtype{gp dt solid}
\draw[gp path] (10.111,7.076)--(11.027,7.076);
\draw[gp path] (0.000,7.229)--(0.781,7.303)--(2.344,6.447)--(3.906,5.084)--(5.468,3.887)%
  --(7.031,2.979)--(8.593,2.398)--(10.155,2.364)--(11.718,2.165)--(12.499,2.068);
\gpcolor{color=gp lt color border}
\gpsetlinewidth{1.00}
\draw[gp path] (0.000,8.381)--(0.000,0.985)--(12.499,0.985)--(12.499,8.381)--cycle;
\gpdefrectangularnode{gp plot 1}{\pgfpoint{0.000cm}{0.985cm}}{\pgfpoint{12.499cm}{8.381cm}}
\end{tikzpicture}
  }
  \caption{The difference of averaged SMBH mass due to the seed BH mass at $z \sim 2$
           (dash-doted line), $z \sim 1$ (doted line), and $z \sim 0$ (solid line)
           as a function of their bulge stellar mass 
           with the \nugc~-H2 simulation.
           The difference becomes smaller at higher redshift.}
  \label{fig:redshift_evolution}
\end{figure}

We explored whether the measurements of the $M_\mathrm{BH}$ -- $M_\mathrm{bulge}$ relation
at higher redshifts help to obtain further constraints on the mass of seed BHs.
Fig. \ref{fig:redshift_evolution} depicts the ratio of the average BH masses in the light seed model
($\equiv\langle M_\mathrm{BH}\rangle_{\scriptscriptstyle 3}$) 
and those in the massive seed model ($\equiv\langle M_\mathrm{BH}\rangle_{\scriptscriptstyle 5}$),
as a function of bulge masses obtained from the \nugc~-H2 simulation.
The difference in the seed mass clearly appears in galaxies 
with bulge mass below $3\times10^{9} M_\odot$ at $z \sim0,1$ and $2$.
We find that the difference due to the seed mass becomes smaller at higher redshift
for a given $M_\mathrm{bulge}$.
Therefore observations of less massive bulges at $z \sim 0$ are
more powerful than at higher redshifts
for constraining the mass distribution of seed black holes.

\begin{figure}
  \centering
  \includegraphics[width=\columnwidth]{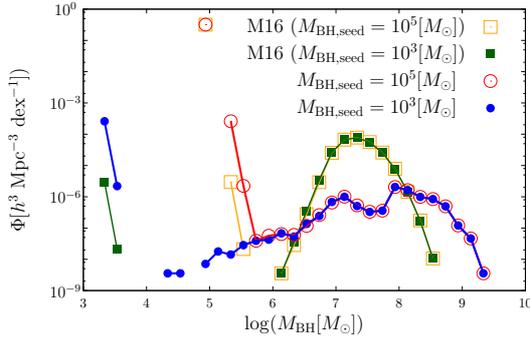}
  \caption{SMBH mass function at $z \sim 6$
           based on the largest cosmological $N$-body simulation (\nugc~-L).
           Orange open and green filled squares with lines 
           show the results of \protect\citetalias{nu2gc},
           red open and blue filled circles with lines show the results
           of light and massive seed models, respectively.
           Even at $z \sim 6$, SMBH mass function
           does not depend on the mass of the seed BHs in the range of
           $M_\mathrm{BH} \gtrsim 10^{5.8} M_\odot$.
           }
  \label{fig:z6}
\end{figure}

Next we investigated SMBH mass functions at $z \sim 6$ 
by using the \nugc~-L simulation.
Fig.~\ref{fig:z6} shows the results 
of the light and massive seed models
(blue and red circles, respectively).
The SMBH mass function at $z \sim 6$ obtained from our model
is nearly consistent with the estimation of \cite{Willott10} in the range of $M_\mathrm{BH} \gtrsim 10^7 M_\odot$.
We find that an SMBH mass function at $z \sim 6$
in the range of $M_\mathrm{BH} \gtrsim 10^{5.8} M_\odot$
does not depend on the mass of the seed BHs
due to the large amount of cold gas
that gets accreted by the BHs.
This is true even when we employ the \citetalias{nu2gc} model,
in which a major merger is the only trigger for
a starburst and gas fueling to a BH.
As mentioned in Sec. \ref{sec:spheroid}, a smaller number of SMBHs, however,
 form at $z \sim 6$ in the \citetalias{nu2gc} model. 
Massive BHs have also grown efficiently by gas accretion
triggered by the disc instability as suggested by \cite{Bower06}.

\section*{Acknowledgements}
We appreciate the detailed review and useful suggestions
by the anonymous referee, which have improved our paper.
We also appreciate the English proofreading by E. J. Tasker,
helpful comments by A. Graham and R. Valiante.
We would like to thank C. Lacey for helpful comments for \nugc.
T.K. was supported in part by an University Research
Support Grant from the NAOJ.
T.O. was financially supported by 
JSPS Grant-in-Aid for Young Scientists (B: 24740112).
R.M. was supported in part by MEXT KAKENHI (15H05896).
T.I. has been supported by MEXT HPCI STRATEGIC PROGRAM and MEXT/JSPS KAKENHI
(15K12031) and by Yamada Science Foundation.
M.N. was  supported by the Grant-in-Aid (25287049) from the MEXT
of Japan.




\bibliographystyle{mnras}
\bibliography{Astrophysics} 








\bsp	
\label{lastpage}
\end{document}